# Electric-field Control of Magnetism with Emergent Topological Hall Effect in SrRuO$_3$ through Proton Evolution

Zhuolu Li[1,†], Shengchun Shen[1,†], Zijun Tian[2,†], Kyle Hwangbo[3,†], Meng Wang[1], Yujia Wang[1], F. Michael Bartram[3], Liqun He[3], Yingjie Lyu[1], Yongqi Dong[4,5,6], Gang Wan[5], Haobo Li[1], Nianpeng Lu[1], Hua Zhou[4], Elke Arenholz[7], Qing He[8], Luyi Yang[1,3,*], Weidong Luo[2,9*] and Pu Yu[1,10,11, *]

[1]*State Key Laboratory of Low Dimensional Quantum Physics and Department of Physics, Tsinghua University, Beijing 100084, China*
[2]*Key Laboratory of Artificial Structures and Quantum Control, School of Physics and Astronomy and Institute of Natural Sciences, Shanghai Jiao Tong University, Shanghai 200240, China*
[3]*Department of Physics, University of Toronto, Toronto, Ontario M5S 1A7, Canada*
[4]*Advanced Photon Source, Argonne National Lab, Argonne, IL 60439, USA*
[5]*Materials Science Division, Argonne National Lab, Argonne, IL 60439, USA*
[6]*National Synchrotron Radiation Laboratory, University of Science and Technology of China, Hefei, Anhui 230026, China*
[7]*Advanced Light Source, Lawrence Berkeley National Laboratory, Berkeley, California 94720, USA*
[8]*Department of Physics, Durham University, Durham DH13LE, United Kingdom*
[9]*Collaborative Innovation Center of Advanced Microstructures, Nanjing 210093, China*
[10]*Collaborative Innovation Center of Quantum Matter, Beijing, China*
[11]*RIKEN Center for Emergent Matter Science (CEMS), Wako 351-198, Japan*
[†]*These authors contributed equally to this work.*
*Email: physics.yangluyi@gmail.com, wdluo@sjtu.edu.cn and yupu@mail.tsinghua.edu.cn

**Ionic substitution forms an essential pathway to manipulate the carrier density and crystalline symmetry of materials via ion-lattice-electron coupling, leading to a rich spectrum of electronic states in strongly correlated systems. Using the ferromagnetic metal SrRuO$_3$ as a model system, we demonstrate an efficient and reversible control of both carrier density and crystalline symmetry through the ionic liquid gating induced protonation. The insertion of protons electron-dopes SrRuO$_3$, leading to an exotic ferromagnetic to paramagnetic phase transition along with the increase of proton concentration. Intriguingly, we observe an emergent topological Hall effect at the boundary of the phase transition as the consequence of the newly-established Dzyaloshinskii-Moriya interaction owing to the breaking of inversion symmetry in protonated SrRuO$_3$ with the proton compositional film-depth gradient. We envision that electric-field controlled protonation opens a novel strategy to design material functionalities.**

**One Sentence Summary:** We report the electric-field controlled magnetism with emergent topological Hall effect in SrRuO$_3$ by proton evolution.

In transition metal oxides, changing the charge carrier concentration can give rise to exotic electronic and magnetic properties, such as the high temperature superconductivity, colossal magnetoresistance, metal-insulator transition, ferromagnetic to paramagnetic transition, etc. (*1-4*) Modifying the crystalline symmetry adds another possibility to engineer the electronic and magnetic states, and the breaking of inversion symmetry in materials can lead to the emergence of ferroelectricity, exotic polar metals (*5-7*) and magnetic skyrmions (*8-16*). To control the carrier density and inversion symmetry, electrostatic charge modulation methods through dielectrics, ferroelectrics and ionic liquids have been widely employed, demonstrating great reversible tunability of material properties (*17-19*). However, an intrinsic limitation of these electrostatic approaches is that they are only effective for materials with thicknesses of a few nanometers, due to the short screening length of the charge carriers. On the other hand, electric field induced ionic evolution recently demonstrates a very good tunability of bulk compounds (*20-22*). Here we demonstrate an efficient and reversible tunability of both the carrier density and the crystalline symmetry within $SrRuO_3$ thick film through ionic liquid gating (ILG) induced protonation. We observe the transition from a ferromagnetic metal to a paramagnetic metal upon protonation, as well as the emergence of a topological Hall effect (THE) due to the breaking of inversion symmetry through the proton concentration gradient.

Perovskite ruthenates have been drawing considerable attention in the last several decades because of their fascinating magnetic and transport properties (*23, 24*). Among them, $SrRuO_3$ demonstrates a unique moderately correlated ferromagnetic metallic state, in which the chemical substitution of Ru cation with Rh can lead to electron doping, resulting in a change from a ferromagnetic to a paramagnetic state (*4*). More interestingly, a THE emerges in ultrathin $SrRuO_3$ films, which is attributed to the breaking of inversion symmetry by the presence of inequivalent interfaces (*14-16*). Therefore, the distinct and rich magnetic transition correlated to the carrier density and inversion symmetry makes $SrRuO_3$ a perfect model system to explore the electronic and magnetic phase diagram through electric-field controlled protonation.

Our experiments were performed on high quality epitaxial SrRuO$_3$ films (with thicknesses ranging from 20 to 90 nm) grown on SrTiO$_3$ (001) substrates by pulsed laser deposition (**Methods**) (*25*). To explore the tunability of electric-field induced protonation, we first performed an *in-situ* X-ray diffraction (XRD) measurements during the ILG. A positive voltage was applied to drive the positively charged protons into the film. **Figure 1A** shows the gate voltage ($V_G$) dependent $\theta$-$2\theta$ scans around the SrRuO$_3$ (002) peak, in which the SrRuO$_3$ (002) peak shows no obvious shift with $V_G$ up to 1.8 V, while a further increase of $V_G$ leads to a clear shift of the peak position to a lower angle (from 45.90° to 44.35°). This result suggests a large out-of-plane lattice expansion up to 3.3% for the SrRuO$_3$ film, which is indeed comparable with our recent result of the pronation induced phase transition from SrCoO$_{2.5}$ to H$_x$SrCoO$_{2.5}$ (*20*). Importantly, the crystalline quality of the film remains unchanged throughout ILG as evinced by the reciprocal space mapping, rocking curves, and reflectivity measurements (**Fig. S1**). Additionally, the phase transformation shows good reversibility. When the gate voltage is removed, the diffraction peak returns nearly to the original position with a slight offset, and afterwards the phase transformation can be reversibly and reproducibly controlled with the application of positive $V_G$ (3.5 V) and zero field (**Fig. 1B**). Notably, the structural transformation possesses a threshold gating voltage (**Fig. S2**).

As our film thickness is much larger than the screening length associated with the electrostatic gating, we can readily single out the ionic (H$^+$ or O$^{2-}$) evolution as the dominant mechanism for the observed structural phase transformations (*25-27*). To identify the ion responsible for the phase transformation, the secondary-ion mass spectrometry (SIMS) was performed, as shown in **Fig. 1C** and **Fig. S3A**. The result shows considerable amounts of hydrogen distributed in the SrRuO$_3$ film associated with a structural phase transformation after being gated at $V_G$ = 3.5 V, but not in a pristine sample or in the film that remains in a pristine state after being gated at 1.5 V. On the other hand, the O$^{18}$ isotopic calibration measurements (*21*) suggest that the oxygen ion evolution is negligible for the gated samples (**Fig. S3B**). Therefore, we can conclude that the ILG induced structural phase transformation is strongly associated with the ILG induced protonation.

Crucially, the presence of positively charged protons are known to lead to electron doping within the materials (*20, 28*). To trace the associated valence state evolution in Ru, we performed *in-situ* hard X-ray absorption experiments near the Ru K-edge for both pristine and protonated (with a gate voltage of 3.5 V) samples, as shown in **Figure 1D**, along with two referenced compounds ($RuO_2$ and Ru metal) (*29*). Clearly, a significant energy shift in the Ru K-edge spectra was observed with respect to that of the pristine state, suggesting the reduction of the Ru valence state due to the electron doping associated with the protonation.

Since the structural phase transformation can be gradually controlled during the ILG, the current study provides a unique opportunity to investigate the evolution of the electronic state of $SrRuO_3$ upon electron doping through protonation. **Figure 2A** shows the temperature dependent resistivity $\rho_{xx}(T)$ for $SrRuO_3$ with different $V_G$ during ILG, in which the thin film remains metallic throughout the gating. However, a careful analysis reveals that a kink feature, which is at ~ 160 K (Curie temperature $T_C$) for pristine samples, is gradually suppressed and eventually disappears (inset of **Fig. 2A**). These results suggest a possible suppression of ferromagnetism during the ILG, as the kink feature is a typical characteristic for ferromagnetism in $SrRuO_3$. This magnetic transition can also be observed in the magnetoresistance (MR = $\rho_{xx}(H)/\rho_{xx}(0)-1$) measurements, as shown in **Figure 2B**. As $V_G$ increases, the negative MR gradually decreases, and more interestingly, with the gating voltage of 2.5 V, we observed a positive parabolic MR, representing a conventional metallic state, instead of the negative butterfly-like MR seen in the pristine ferromagnetic sample.

To clearly investigate the evolution of the ferromagnetic state in $SrRuO_3$ under ILG induced protonation, we measured the magnetic-field dependent Hall resistivity at different values of $V_G$. The pristine $SrRuO_3$ film exhibits a well-defined hysteresis loop attributed to the anomalous Hall effect (AHE) associated with the ferromagnetic state at low temperatures. As $V_G$ increases, the hysteresis loop (at 2 K) is gradually suppressed and eventually turns into a linear response with $V_G$ of 2.5 V, as shown in **Fig. 2C**. **Figure 2D** summarizes the AHE resistivity (extracted at $\mu_0 H = 0$ T) at different temperatures under various $V_G$, in which the anomalous Hall resistivity gradually decreases and eventually disappears with the increase of $V_G$.

The magnetic evolution during ILG was further studied with the *in-situ* magneto-optic Kerr effect (MOKE) measurements, which measures the ac inter-band Hall conductivity and has the same origin as the *intrinsic* AHE (i.e. the anomalous velocity due to the Berry curvature in momentum space (*30*)). Similar to the Hall measurements, as $V_G$ increases the square-like MOKE hysteresis loop is gradually suppressed and eventually disappears (**Figs. 2E and 2F**), indicating that the ferromagnetism is indeed weakened by the ILG induced protonation. Furthermore, the element-specific X-ray magnetic circular dichroism (XMCD) measurements at the Ru $L_{3,2}$ edges clearly show the suppression of ferromagnetism in Ru ions in the protonated sample (**Fig. S4**). Undoubtedly, all these observations provide strong evidence that the protonated $H_xSrRuO_3$ sample undergoes an exotic ferromagnetic to paramagnetic phase transformation with the protonation induced electron modulation. Furthermore, like the structural transformation, the modulation of ferromagnetic state is reversible when cycling $V_G$ (**Fig. S5**). The slight suppression of the AHE signal (as well as the magnetization) after removing the gating voltage compared to the pristine samples (**Figs. S5** and **S6**) should be attributed to the proton residual previously observed in the structural modulation and in the SIMS results (**Fig. 1B, C**).

To shed more light on the protonation induced structural and magnetic phase transformations, we carried out first-principles calculations. The optimized crystalline structure of $HSrRuO_3$ is shown in the inset of **Figure 1A**, in which the proton is bonded with the equatorial oxygen of the Ru octahedral. **Figures 3A and 3B** present the calculated non-spin-polarized band structures for pristine and protonated $HSrRuO_3$ samples, respectively. Clearly, the proton intercalation leads to a significant splitting of the degenerated Ru $t_{2g}$ bands with dramatically modified density of states (DOS). As shown in **Fig. 3C**, although the spin-resolved DOS shows significant splitting of majority (down) and minority (up) bands in pristine SrRuO$_3$, the corresponding DOS in protonated HSrRuO$_3$ shows a nearly equivalent spectral weight, indicating the absence of ferromagnetism in the latter. It has been well established that the metallic ferromagnetism of SrRuO$_3$ can be described within the framework of the Stoner model (*4, 31*), in which the ferromagnetic ground state is favored when $IN_0 > 1$, where *I* and $N_0$ are the so-called Stoner factor and nonmagnetic DOS per spin at the $E_F$ respectively. Clearly, the change of DOS at the $E_F$ would play an important role for the corresponding magnetic states. To trace the magnetic ground state, we have calculated crystalline structures as well as the Stoner factor (and then $IN_0$ value) for a series of protonated H$_x$SrRuO$_3$ with different proton concentrations, as summarized in **Figure 3D**. The results show that with the intercalation of protons, the lattice results in a dramatic expansion, being consistent with the XRD results. More interestingly, the calculations show that with the increasing proton concentration, the value of $IN_0$ gradually decreases. According to the Stoner criterion, a non-magnetic (or paramagnetic) ground state would be favored for the case with $IN_0 < 1$, therefore this theoretical calculation therefore nicely explains our experimental observations of protonation induced ferromagnetic to paramagnetic transition in the SrRuO$_3$ film.

We note that the proton intercalation through ILG is dominated mainly by the diffusion process. With a fine tuning of the gating voltage, we observed a dramatic broadening of the XRD diffraction peak at certain voltage (**Figs. S7A** and **S7B**), suggesting the formation of inhomogeneous protonation along the surface normal as the intermediate state. Indeed, we observed a considerable proton concentration gradient in an *ex-situ* H$_x$SrRuO$_3$ sample (**Fig. 4A**) formed at the boundary of the phase transition. Moreover, an increased second harmonic generation (SHG) signal was also detected in the H$_x$SrRuO$_3$ state as compared to the pristine film (**Figs. 4B** and **S7C**), which can be explained by the facts that the broken inversion symmetry allows the bulk, rather than just the surface, to contribute to the SHG signal. Therefore, the ILG induced protonation provides a novel pathway to control the inversion symmetry within the material, in which the proton concentration gradient leads to a broken inversion symmetry through a built-in polarization field.

Previous studies have reported that the broken inversion symmetry through inequivalent interfaces in ultrathin SrRuO$_3$ films can couple with their intrinsic spin-orbit coupling to produce a strong Dzyaloshinskii-Moriya (DM) interaction, leading to the emergence of an exotic THE effect (*14-16*). Inspired by these observations, we turned our attention to the novel magnetic states in the intermediate protonated sample. A careful examination of the data presented in **Figure 2C** indeed reveals that the AHE result with gating voltages around 1.8 V shows a distinct hump feature (**Fig. 4C**), a hallmark for the THE observed in bulk MnSi (*9, 10*), EuO thin films (*13*) and SrIrO$_3$/SrRuO$_3$ heterostructures (*14-16*). Interestingly, the THE effect is tunable with the gating voltage, as shown in **Fig. S8**.

To quantitatively evaluate the THE, we estimated the topological Hall resistivity $\rho_{YX}^T$ (as shown in **Fig. 4C**) by subtracting the AHE signal through linear fitting of the high field data. With this, we can construct a phase diagram for the topological Hall term $\rho_{YX}^T$ in the *T-H* plane (**Fig. 4D**), showing that the THE clearly exists in a wide range of the *T-H* plane. Moreover, although the sign of the AHE component remains unchanged up to 100 K, the corresponding THE component changes sign with the increase of temperature and both positive and negative THE components are identified at certain temperatures (*e.g.* 10 K), suggesting multiple spin textures with scaler spin chirality are induced at corresponding temperatures. In addition, we confirm that the THE is driven by the magnetization reversal process due to the fact that the peak position of THE ($H_P$) scales nicely with the coercive field $H_C$. We note that in previous studies of ultra-thin SrRuO$_3$ film (~2-3 nm) systems, the THE emerges as the consequence of the enhanced DM interaction as well as reduced ferromagnetism due to the interface effect (*14-16*). In our system, the ILG induced large proton compositional gradient at the boundary of the ferromagnetic to paramagnetic phase transition would lead to a larger DM interaction as well as clearly reduced ferromagnetism, and both favor the emergence of THE.

To summarize, using the SrRuO$_3$ films as the model system, we have demonstrated an electric-field controlled reversible magnetic transformation through the protonation evolution. Moreover, the induced proton concentration gradient can naturally break the inversion symmetry, resulting in the emergence of an exotic topological Hall effect. These results strongly suggest that the electric-field induced protonation opens a new avenue to systematically control the electronic and magnetic phase diagram in strongly correlated complex oxide systems.

**Note added:** During finalizing this work, we learned that an experimental study has demonstrated a ferroelectric control of magnetic skyrmion (i.e. topological Hall effect) in ultra-thin (4-5 unit cells) SrRuO$_3$ heterostructures (*32*).

# Acknowledgements

This study was financially supported by the Basic Science Center Program of NSFC (grant No. 51788104); NSFC (grant No. 11474197, U1632272 and 11521404); the National Basic Research Program of China (grant No. 2015CB921700 and 2016YFA0301004); and the Beijing Advanced Innovation Center for Future Chip (ICFC). First-principles calculations were performed at the HPC of Shanghai Jiao Tong University, China. The optical measurements were carried out at the University of Toronto and were supported by CIFAR Azrieli Global Scholars, Canada Research Chair, NSERC, CFI, ORF and UofT startup funds. This research used resources of the Advanced Photon Source, a U.S. Department of Energy (DOE) Office of Science User Facility operated for the DOE Office of Science by Argonne National Laboratory under Contract No. DE-AC02-06CH11357. This research used resources of the Advanced Light Source, which is a DOE Office of Science User Facility under contract no. DE-AC02-05CH11231.

# Author contributions

P.Y. conceived the project and designed the experiments. Z.L. grew the sample and performed the XRD, SIMS and XPS measurements with helps from M. W., Y. W., H. L. and N. L.. S.S. conducted the transport measurements and analyzed the data. Z.T. performed the first-principles calculations under the supervision of W. L.. K. H. built the MOKE experiment and performed the MOKE measurements with the help from L. H., and F. M. B. set up the SHG experiments and carried out the SHG measurements under the supervision of L. Y.. Y. W., Y. L., E. A. and Q. H. performed the soft X-ray XMCD measurements. Y.D., G.W. and H. Z. performed the *in-situ* XANES measurements. Z.L., S. S. and P. Y. wrote the manuscript, and all authors commented on the manuscript.

## Supplementary Materials:

Materials and Methods

Supplementary Text

Figures S1-S8

References

# Figures and Captions

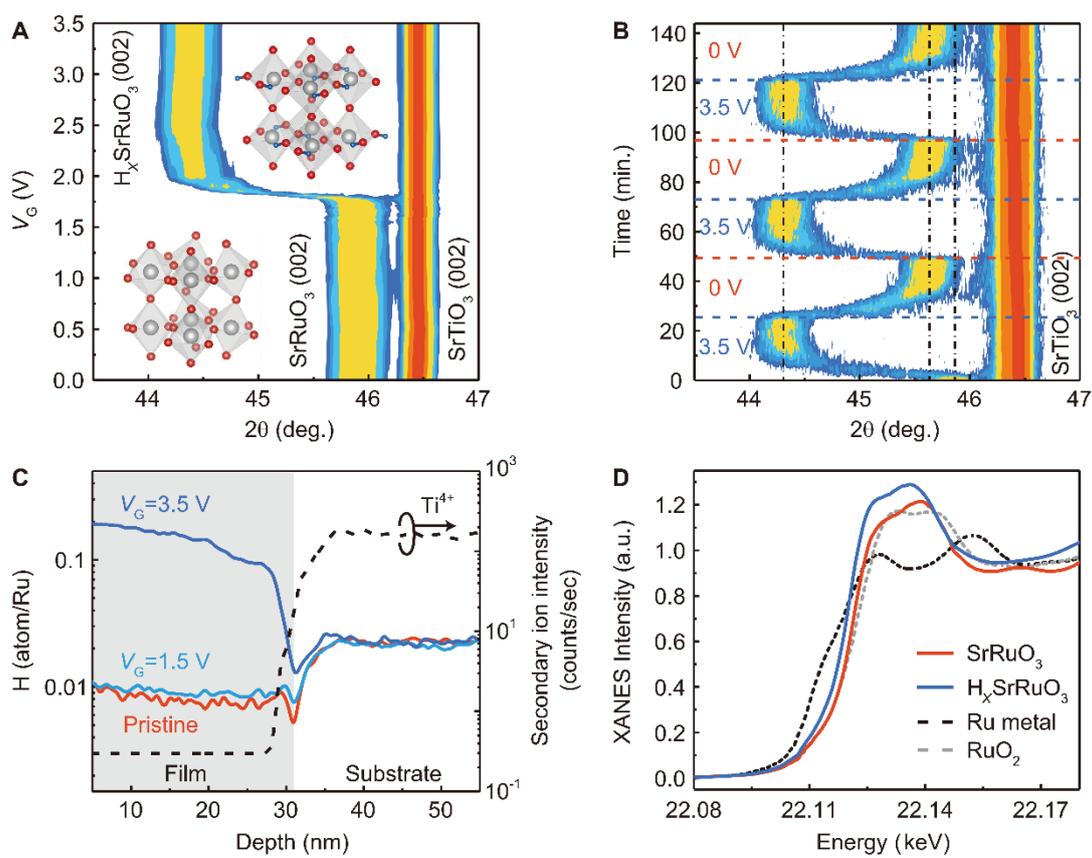

**Fig. 1. Electric-field controlled protonation in SrRuO₃ films.** (**A**) *In-situ* XRD $\theta$-$2\theta$ scans around the SrRuO$_3$ (002) peak as a function of $V_G$. The new phase is denoted as H$_x$SrRuO$_3$. The insets show the calculated crystal structures of SrRuO$_3$ and HSrRuO$_3$, respectively, where the blue balls represent hydrogen atoms. (**B**) *In-situ* XRD $\theta$-$2\theta$ scans around the SrRuO$_3$ (002) peak as $V_G$ cycled between 3.5 V and 0 V, indicating the reversibility of the structural phase transformations. The black dotted lines indicate the peak positions, and the blue and red dash line denotes the change of $V_G$. (**C**) Hydrogen distribution profiles within both pristine SrRuO$_3$ films (red solid line) and gated (with $V_G$ = 3.5 V or 1.5 V) SrRuO$_3$ films, as measured *ex-situ* by SIMS. The Ti$^{4+}$ signature was used as a marker to define the interface between film and substrate. (**D**) *In-situ* XANES spectra at Ru K-edge for pristine SrRuO$_3$ (orange solid line) and protonated H$_x$SrRuO$_3$ (blue solid line) at $V_G$ = 3.5 V. The XANES spectra at Ru K-edge for Ru metal (black dash line) and RuO$_2$ (gray dash line) are shown as references.

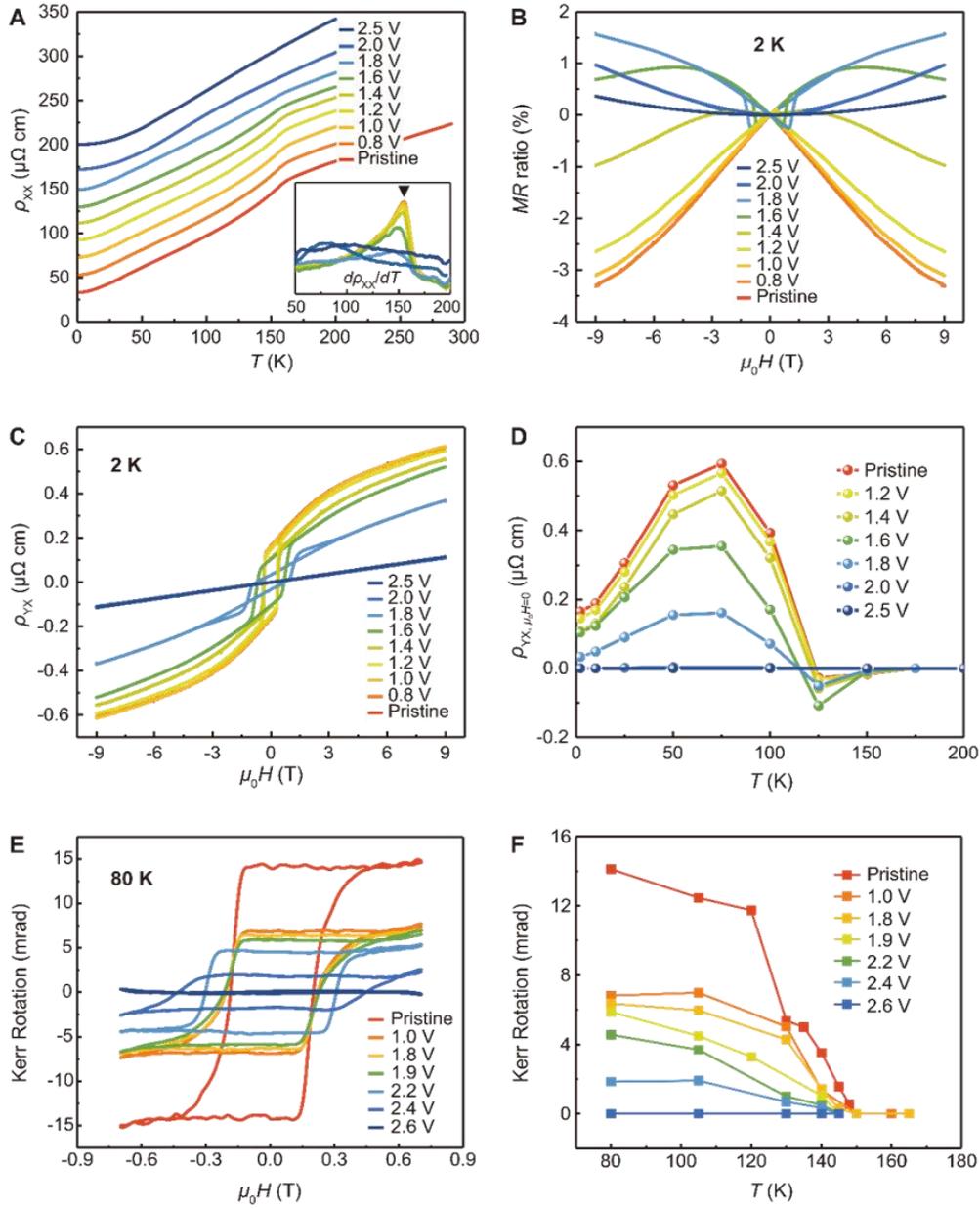

**Fig. 2. Electric-field controlled magnetic evolution via protonation.** (**A**) Temperature dependent transverse resistivity $\rho_{xx}$ at different $V_G$. The inset shows the corresponding differentiate resistivity $d\rho_{xx}/dT$ at different $V_G$. A vertical offset of 20 μΩ cm is applied for each curve for clarity. (**B**) Magnetic field dependent magnetoresistance (MR) measured at 2 K with different $V_G$. (**C**) Magnetic field dependent Hall resistivity measured at 2 K with different $V_G$. (**D**) Temperature dependent anomalous Hall resistivity obtained at $\mu_0 H = 0$ T with different $V_G$. (**E**) Kerr rotation vs. magnetic field results measured at 80 K with different $V_G$. (**F**) Kerr rotation as a function of temperature obtained at $\mu_0 H = 0$ T with different $V_G$. The slightly varied threshold gate voltages among transport, MOKE and XRD measurements are attributed to the different device configurations.

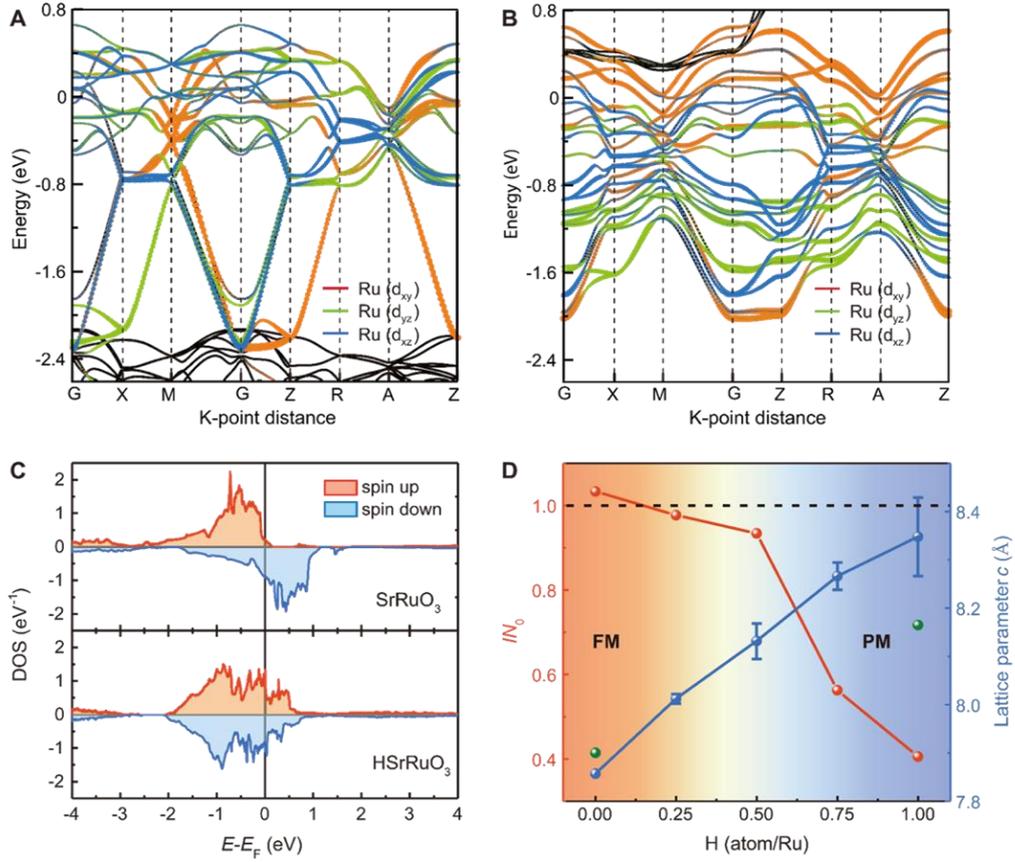

**Fig. 3. Mechanism for the protonation induced magnetic phase transition.** Calculated electronic band structures for (**A**) pristine SrRuO$_3$ and (**B**) HSrRuO$_3$ with nonmagnetic General Gradient Approximation (GGA) calculations. (**C**) Spin-resolved density of states for pristine SrRuO$_3$ (upper panel) and protonated HSrRuO$_3$ (bottom panel) calculated by GGA. (**D**) Calculated Stoner criterion parameter $IN_0$ and *c*-axis lattice parameter as a function of hydrogen concentration. The green points are the experimental lattice parameters (doubled of the pseudocubic lattice constant) obtained from the *in-situ* XRD measurements. Following the Stoner criterion, when $IN_0$ becomes smaller than 1 with the increase of hydrogen concentration, the ferromagnetic (FM) SrRuO$_3$ transits into a paramagnetic (PM) metal.

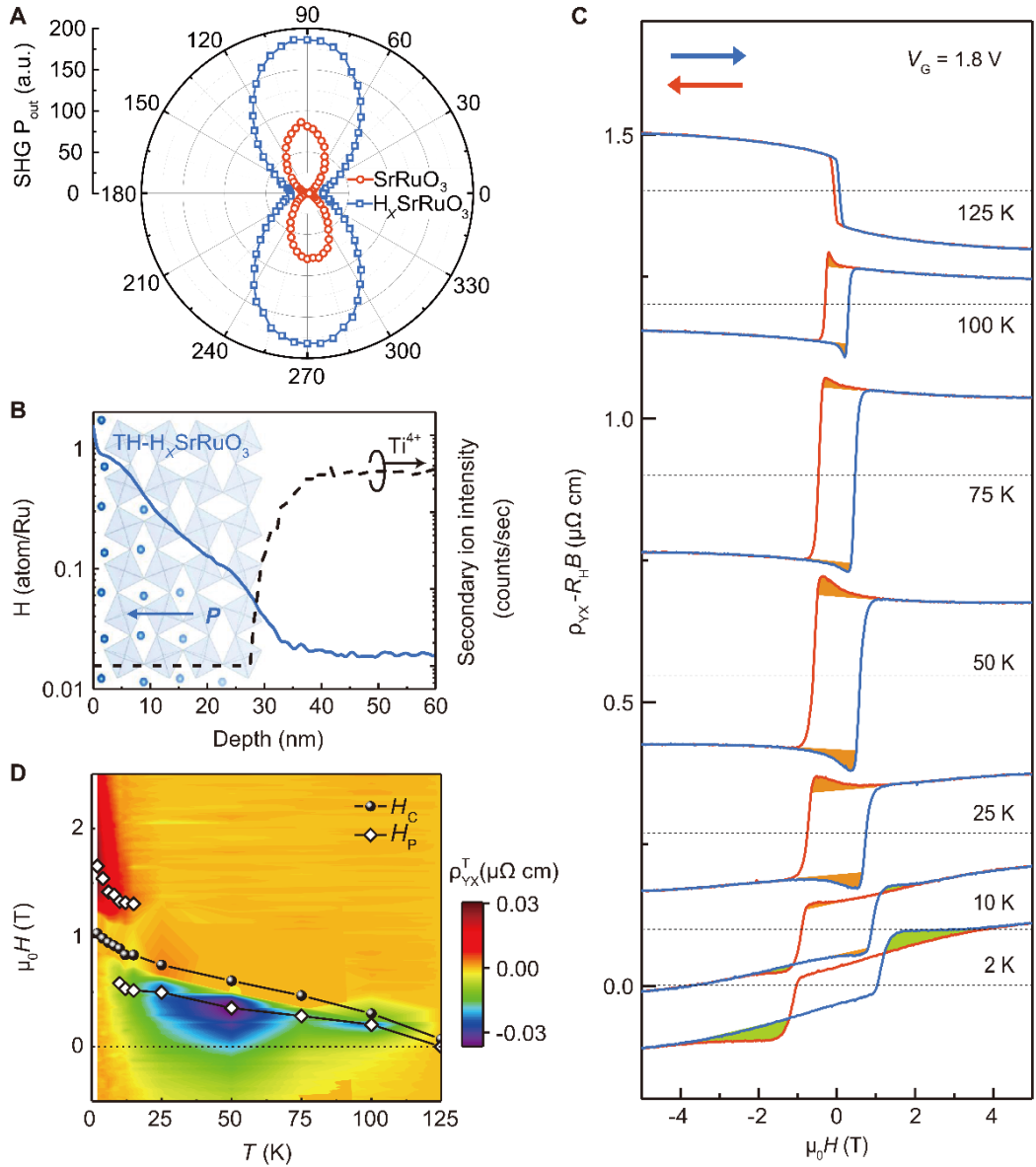

**Fig. 4. Emergence of topological Hall effect in H$_x$SrRuO$_3$.** (**A**) *Ex-situ* measured proton distribution profile in a sample (32 nm) with topological Hall effect. The inset illustrates the gradient of the proton concentration in the gated H$_x$SrRuO$_3$, leading to the formation of a net polarization along the out-of-plane direction and thus breaks the inversion symmetry. (**B**) The p-polarized SHG intensity as a function of the polarization direction of the incident light (0 corresponds to s-polarization) for both pristine SrRuO$_3$ and gated H$_x$SrRuO$_3$ films. The weaker SHG in the pristine film is due to surface contributions, while the enhanced SHG intensity of the H$_x$SrRuO$_3$ state suggests the breaking of inversion symmetry in bulk. (**C**) Magnetic field dependent Hall resistivity for H$_x$SrRuO$_3$ gated with $V_G$ = 1.8 V at different temperatures. The blue and red arrows denote the field sweeping direction. Ordinary Hall term is subtracted through the linear fitting of $R_H B$ at higher magnetic fields. An offset is applied per curve for clarity, while the dotted lines denote the center of the hysteresis loops. The estimated topological Hall resistivity with different signs is marked different colors. (**D**) Color map of estimated topological hall resistivity ($\rho_{YX}^T$) and characteristic fields ($H_C$ and $H_P$) obtained at $H_x SrRuO_3$ gated with $V_G = 1.8$ V. $H_C$ (black filled symbol) represents the coercive field and the $H_P$ (white open diamond) denotes the field where the topological Hall resistivity reaches its maximum.

# Supplementary Materials for

**Electric-field Control of Magnetism with Emergent Topological Hall Effect in SrRuO$_3$ through Proton Evolution**

**This PDF file includes:**

  Materials and Methods

  Supplementary Text

  Figs. S1 to S8

  References

## Materials and Methods

**Film growth and X-ray diffraction measurements:** Epitaxial SrRuO$_3$ films were grown on (001) SrTiO$_3$ and (001) LaAlO$_3$ substrates (only for *in-situ* XANES measurements) by pulsed laser deposition (KrF, λ = 248 nm). All films were deposited at identical conditions with a substrate temperature of 700 ℃ and an oxygen partial pressure of 100 mTorr. The energy density of laser was fixed at 2 J/cm$^2$ with a growth rate of 1.7 nm/min. To minimize the oxygen vacancy concentration, samples were post-annealed at growth temperature for 15 min and then cooled down to room temperature at a cooling rate of 10 ℃ per minute in an atmosphere of oxygen pressure. For *in-situ* XRD measurements, Au electrodes were sputtered at edges of films, and a slice of Pt was selected as the gated electrode. Both sample and Pt were placed in a quartz bowl, and then the whole gating device was put on the XRD sample-holder. The covered ionic liquid (IL) was carefully controlled so as to get strong enough diffracted X-ray signal. The *in-situ* θ-2θ and reciprocal space mapping (RSM) measurements were performed with a high-resolution diffractometer (Smartlab, Rigaku) using monochromatic Cu Kα1 radiation (λ = 1.5406 Å) at room temperature. The $V_G$ dependent θ-2θ scans were measured with the same time interval for each curve.

***In-situ* electrical transport and *ex-situ* magnetic measurements:** The transport measurements were performed in a PPMS setup (Quantum Design DynaCool system, 9 T) equipped with lock-in amplifiers (Model LI 5640, NF Corporation). Hall-bar structures (220 μm × 60 μm) were fabricated by standard lithography and Au/Ti was sputtered as electrodes. The device was placed in a quartz bowl covered entirely with IL and a slice of Pt was used as the gate electrode. To exclude the offsets for Hall resistivity and longitudinal resistivity due to misalignment of the contacts, the Hall resistivity $\rho_{xy}$ and longitudinal resistivity $\rho_{xx}$ were calculated from the raw data as:

$$\rho_{xy} = t[V_H(+H \rightarrow -H) - V_H(-H \rightarrow +H)]/2I,$$

$$\rho_{xx} = Wt[V_R(+H \rightarrow -H) + V_R(-H \rightarrow +H)]/2IL,$$

where *t*, *W* (60 μm), *L* (220 μm) are the thickness, width and length of the device respectively and *I* is the excitation current. $V_G$ was applied between the gate electrode (Pt) and the SRO film. For each $V_G$, the measured conditions were the same. The $V_G$ was changed at 290 K and then dwelled for 25 minutes for each cycle. Magnetic characterizations, including the temperature-dependent magnetization (*M–T*) and magnetic hysteresis loops (*M–H*), were carried out with a 7 T SQUID magnetometer (Quantum Design). Both electrical and magnetic measurements were done within the vacuum cryostats.

***In-situ* magnetic-optical Kerr effect (MOKE) measurements:** The MOKE measurements were carried out using a narrowband continuous-wave diode laser at 895 nm wavelength. The sample was mounted on the cold finger of a small optical cryostat (3-300 K). One edge of the sample was painted with the silver conductive adhesive to act as the bottom electrode. Platinum wire surrounding the sample was used as the gate electrode. The sample was then covered by a droplet of ionic liquid and was further covered by a coverslip to obtain the flat surface. Magnetic field (up to 0.7 T) generated using an external electromagnet was applied perpendicular to the sample plane (Faraday geometry). The light (250 μW) was linearly polarized and focused weakly into a 50 μm spot on the sample surface at normal incidence. The out-of-plane magnetization of the sample was then detected by measuring the optical Kerr rotation ($\theta_K$) of the reflected light. The light intensity was modulated with a mechanical chopper at 1.2 kHz to facilitate a lock-in detection. The Kerr rotation signals as a function of the applied magnetic field were measured with a standard optical bridge arrangement using balanced photodiodes.

**Second harmonic generation (SHG) measurements:** To measure the SHG signal, a Ti: Sapphire pulsed laser (800 nm wavelength, 200 fs pulse duration, 76 MHz repetition rate, average power 40 mW) was linearly polarized and focused onto the sample at a 45° incident angle using a 0.35 NA microscope objective. The reflected beam was collimated and passed through a linear polarizer and two 450 nm short pass filters to remove all laser light at 800 nm. The second harmonic (400 nm) light was then detected using a photomultiplier tube connected to a current preamplifier. The signal was modulated at approximately 3 kHz using an optical chopper and detected with a lock-in amplifier to improve the signal to noise ratio. For each setting of the final linear polarizer, the SHG power was measured as a function of the initial polarization angle of the laser. The sample was also mounted on a manual rotation stage to allow different orientations of the sample with respect to the plane of incidence to be measured.

**Compositional analysis through the phase transformation:** The secondary ion mass spectroscopy (SIMS) measurements for H, $^{18}$O and deuterium (D or $^2$H) were carried out using a TOF-SIMS 5-10 instrument (IONTOF GmbH). The sputtering area was 250 μm × 250 μm and the detecting area was 50 μm × 50 μm in order to avoid the disturbance from the crater edge. The interface position was indexed by measuring the Ti element from the SrTiO$_3$ substrate. All samples were measured with the same conditions. The concentration of H was estimated by profiling proton-implanted SiO$_2$ with known hydrogen dosage.

*In-situ* **X-ray absorption near-edge structure (XANES) measurement:** XANES studies were performed at the bending magnet beamline, 12-BM-B, at the Advanced Photon Source, Argonne National Laboratory. The linear polarized X-rays after the Si (111) monochromator with resolution $\delta E/E = 1.4 \times 10^{-4}$ has a total flux of $2 \times 10^{11}$ photons/s. The absorption spectra were collected by the fluorescence mode with the samples mounted in a custom-designed X-ray cell allowing *in-situ* electrochemical control of gating bias applied during the ILG. A 13-element Ge drift detector (Canberra) was used to measure the fluorescence yield. Glancing incidence geometry (e.g. > 4-5 times of the substrate critical angle) was adopted to cover the signal contributed by the whole SrRuO$_3$ film as well as to minimize the elastic scattering background. A Ru metal foil was used as an online check of the monochromator energy calibration. The originally collected XANES spectra were normalized by fitting the pre-edge to zero and the post-edge to 1 using Ifeffit performed by the software Athena.

**Soft X-ray absorption spectra and X-ray magnetic circular dichroism measurements:** Soft X-ray absorption spectra (XAS) and corresponding X-ray magnetic circular dichroism (XMCD) measurements at Ru M-edge were measured at beamline 4.0.2 of Advanced Light Source with total electron yield (TEY) mode. The measurements were done at 10 K under high vacuum (around $10^{-8}$ Torr), and the incident circularly polarized (90 %) X-ray was perpendicular to the film surface, with a magnetic field of 4 T applied along the beam incident direction. The XAS spectra were deduced form the average of positive ($\mu^+$) and negative ($\mu^-$) magnetic field contributions, and the XMCD spectra were taken as the difference between two magnetic field contributions.

**First-Principles Calculations**: First-principles density-functional theory (DFT) *(33, 34)* calculations with the generalized-gradient approximation (GGA) *(35)* and the projector-augmented wave (PAW) method *(36)* were performed using the Vienna Ab-initio Simulations Package (VASP) *(37)*. The calculations were performed using a plane-wave energy cutoff of 400 eV, and a Gamma mesh of $4 \times 4 \times 4$ k-points and $2 \times 2 \times 2$ k-points were used for the pristine orthorhombic $SrRuO_3$ and the protonated superstructure, respectively. The lattice constants were fixed to the experimental result during the calculations for the pristine orthorhombic $SrRuO_3$, while in the calculations for the protonated superstructure, the in-plane lattice constant was fixed to 3.905Å (lattice constant of $SrTiO_3$), and the out-of-plane lattice constant was allowed to relax. In both cases, the atomic positions were fully optimized in all the magnetic orders until the change of the total energy is smaller than $10^{-5}$ eV.

## Supplementary Text

### Hydrogen intercalation into the $SrRuO_3$ films during the ILG

Both electrostatic charge accumulation (*26*) or oxygen ion evolution (*27*) have been reported in the previous studies of ILG on $SrRuO_3$ thin films. While in our case, the observations of both structural transformations and reduction of Ru valence state in bulk strongly suggest that the effect is not dominated by the surface electrostatic charge modulation. To trace the possible oxygen evolution during ILG, we also performed SIMS measurements on $SrRuO_3$ films dwelled in pure $O^{18}$ atmosphere during the ILG, where any oxygen evolution occurred during the ILG would result in the accumulation of $O^{18}$ within the film. However, the SIMS measurements reveal that the $O^{18}$ depth profile signals are almost identical and negligible for both pristine and gated $SrRuO_3$ films (**Fig. S3B**). Therefore, this result indicates the oxygen evolution is not essential either during ILG in our case. Thus, we can conclude that the hydrogen intercalation into SrRuO$_3$ films during ILG operates as the dominated mechanism for the observed crystalline structural phase transformation and reduction of Ru valence state across the whole film. With this, we can assign the chemical formula for the new protonated phase as H$_x$SrRuO$_3$. In order to trace the origin of the hydrogen in the H$_x$SrRuO$_3$, we doped the employed ionic liquid with heavy water (D$_2$O). Consequently, significant deuterium (D) signal was observed in the gated SrRuO$_3$ film (**Fig. S3A**), which provides solid evidence to attribute the origin of hydrogen in the gated SrRuO$_3$ films to the water residual (typically around several hundred ppm) inside the ionic liquid (*20, 28*).

**Topological Hall effect in hydrogenated SrRuO$_3$**

With the observation of THE, we would expect the emergence of a nontrivial magnetic skyrmion spin texture in protonated SrRuO$_3$. In the skyrmion systems, the magnitude of the topological hall resistivity is related to the emergent magnetic field derived from the real-space Berry phase. Therefore, we can roughly estimate the skyrmion density with the calculated fictitious magnetic field. A single skyrmion can be regarded as a flux quantum. Hence, the topological Hall resistivity ($\rho_{YX}^T$) induced by skyrmion can be expressed by the formula (*14*) $\rho_{YX}^T = PR_0 n_{sk} \phi_0$, where *P* denotes the spin polarization in SrRuO$_3$, $R_0$ is the ordinary Hall coefficient, $n_{sk}$ is skyrmion density and $\phi_0 = h/e$ is the magnetic flux quantum. Based on tunneling MR measurement in a SrRuO$_3$ junction, the SrRuO$_3$ spin polarization was reported around -9.5% (*38*). Using the maximum estimated topological Hall resistivity (around 0.03 μΩ-cm, **Fig. 4C**), the skyrmion density $n_{sk}$ can be deduced around $2.6 \times 10^{15}$/m$^2$. In this case, the estimated skyrmion size is $n_{sk}^{-1/2} \sim$ 20 nm, which is reasonable compared with the typical DM interaction induced skyrmion size (range from 5 nm to 100 nm) in other system (*12*). We note that since the film was totally immersed into the IL, we are not able to directly probe the skyrmion by the well-employed magnetic force microscopy or Lorentz transmission electron microscopy methods.

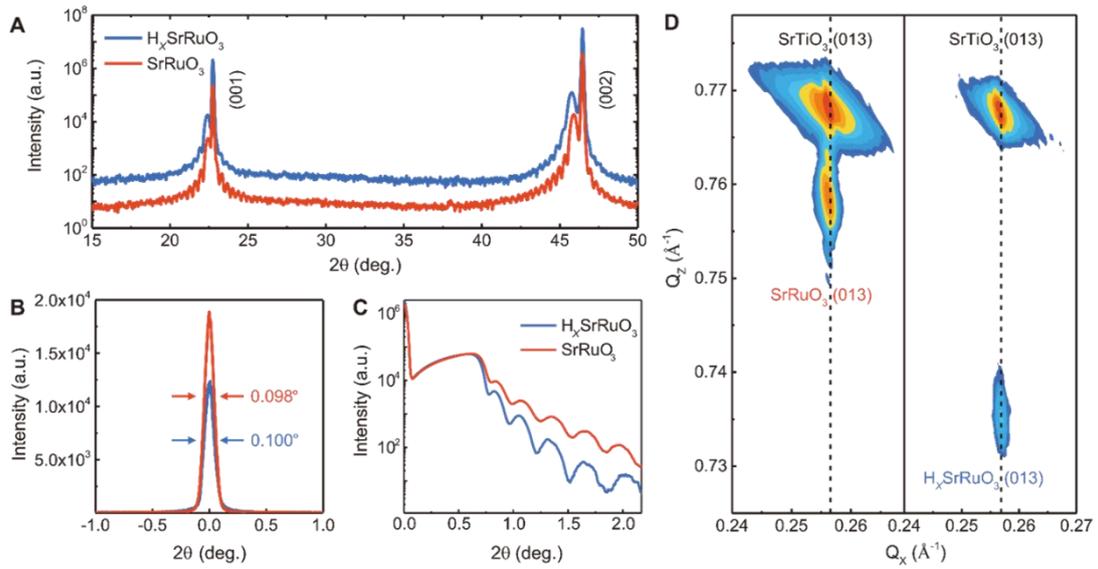

**Fig. S1. Comparison of the crystalline structures of pristine and protonated SrRuO₃ films.** **(A)** XRD $\theta$-$2\theta$ scans of pristine (red, 26 nm) and gated (blue) SrRuO$_3$ films, respectively. **(B)** Rocking curves around (002) peaks of both two phases. The full width at half maximum (FWHM) of pristine and gated films are almost identical. **(C)** X-ray reflectivity curves of these two phases. The protonated samples used for the ex-situ XRD measurements shown in (**A, B, C**) were obtained with ILG at $V_G$ = 3.5 V. **(D)** Reciprocal space mapping (RSM) results around (013) peaks of STO substrates for both pristine and *in-situ* protonated (at $V_G$ = 3.5 V) films.

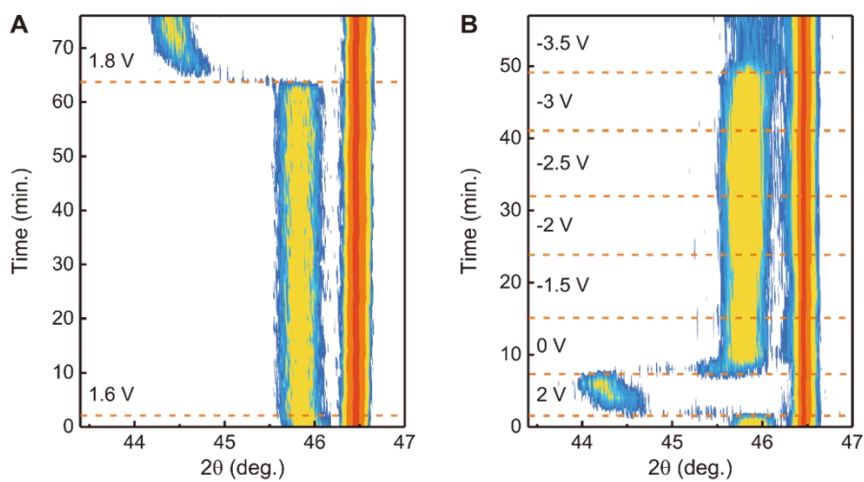

**Fig. S2.** *In-situ* **XRD measurements during the ILG induced phase transformation.** (**A**) Gating duration dependent *In-situ* θ-2θ scans around (002) peak of films as fixed $V_G$ of 1.6 and 1.8 V. (**B**) *In-situ* XRD θ-2θ scans around (002) peak in a wide range of $V_G$ from +2 V to -3.5 V. The protonated phase quickly returns back to a lightly protonated phase under the gating voltage of 0 V and then remains stable with the gating voltage up to -3 V, above which the film will be damaged by the ILG process.

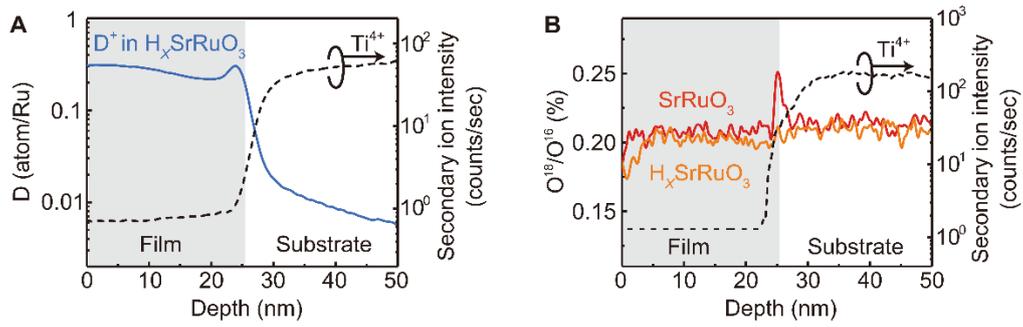

**Fig. S3. Extended compositional analysis of the gated SrRuO₃ films.** (**A**) Depth profiles of the deuterium ion (D$^+$) in gated ($V_G$ of 3.5 V) SrRuO$_3$ film with the heavy water doped IL. (**B**) Depth profile of O$^{18}$/O$^{16}$ ratio in pristine SrRuO$_3$ film and the gated film relaxed in O$^{18}$ atmosphere. In both (**A**) and (**B**), the Ti$^{4+}$ signal was used to define the interface position.

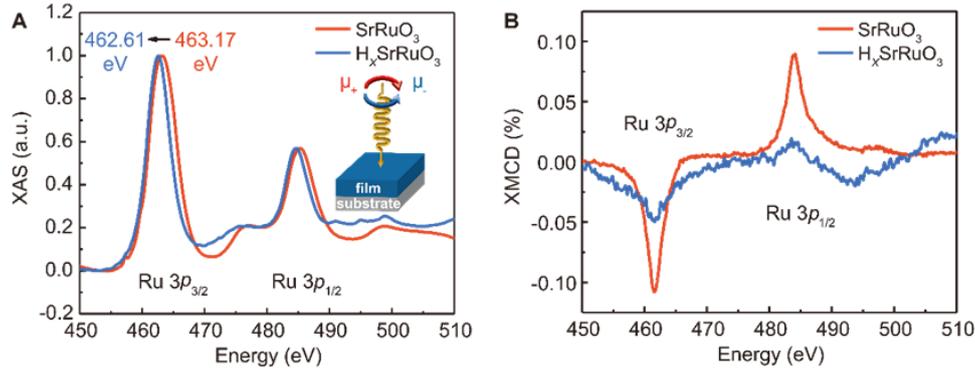

**Fig. S4. X-ray absorption spectra for pristine and protonated SrRuO₃ films. (A)** XAS spectra of pristine (red) and *ex-situ* gated (blue) SrRuO3 films. The shift of peak positions for Ru M$_3$ and M$_2$ peaks to the lower energy indicates the reduced Ru valence state in the gated sample, which is consistent with our scenario of electron doping with protonation. The inset shows the experimental configuration, in which the incident µ⁺ (right circularly) and µ⁻ (left circularly) polarized X-rays are perpendicular to the film surface. **(B)** Comparison of x-ray magnetic circular dichroism (XMCD) results for both two phases. An obvious decrease in the XMCD intensity suggests a suppression of ferromagnetic contribution from Ru element. We note that the soft x-ray absorption technique is a surface technique, which probes mainly the top 5 to 10 nm of the films. While in our gated sample, although the majority portion of proton is released from the film along with the structural relaxation, there is still large amount of proton concentration at the top surface, which can nicely explain the suppressed XMCD signal as the consequence of the reduced magnetism in the protonated SrRuO$_3$.

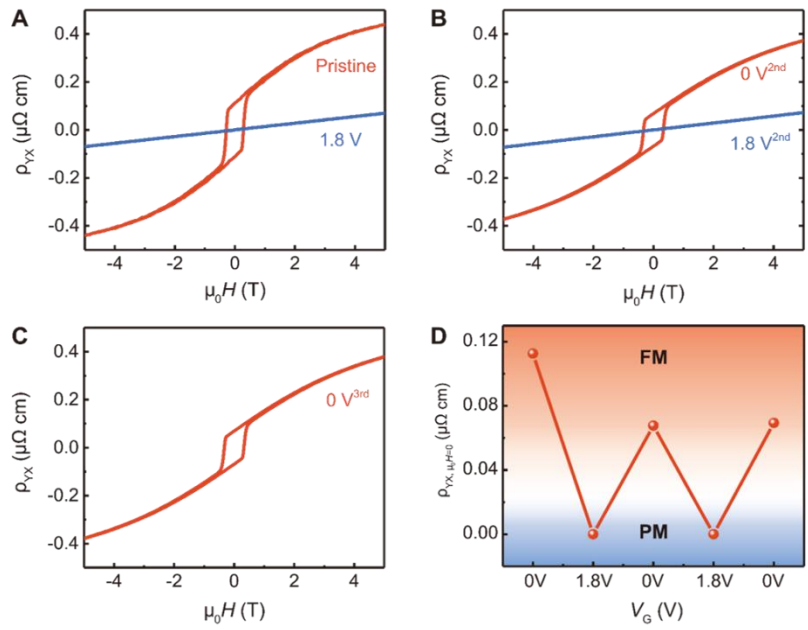

**Fig. S5. Reversible control of magnetism in SrRuO₃ during the ILG.** (A-C) Magnetic field dependent Hall resistivity at 2 K as the gating voltage $V_G$ was cycled between 0 V and 1.8 V in another sample. **(D)** The reversible ferromagnetic phase transition along with cycled $V_G$.

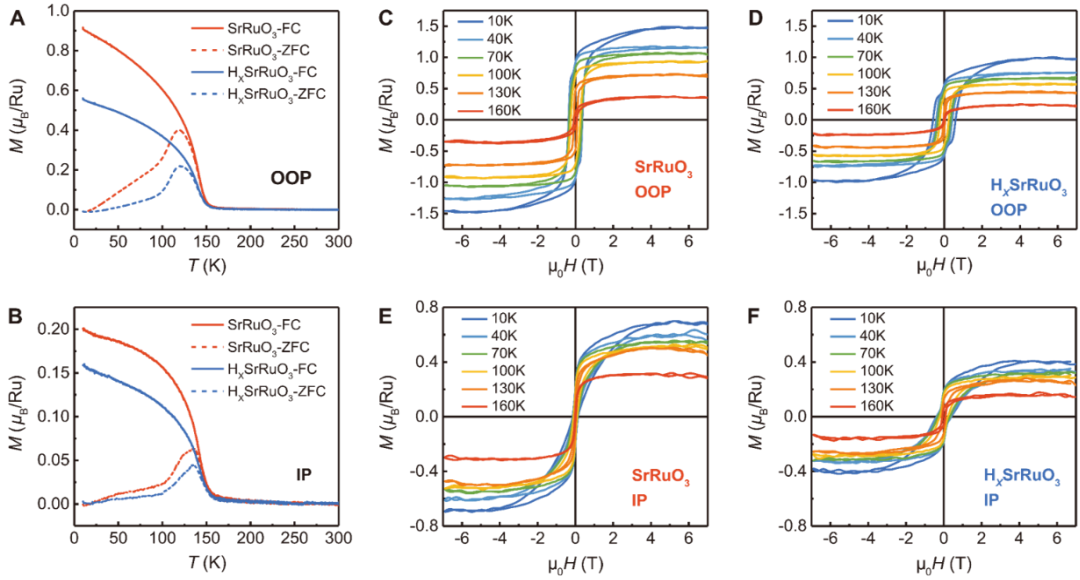

**Fig. S6. Macroscopic magnetic measurements of pristine and post-gated SrRuO₃ films.** Temperature dependent (**A**) out-of-plane (OOP) and (**B**) in-plane (IP) magnetization with field cooling and zero field cooling processes for both pristine and gated samples. (**C, D**) Out-of-plane and (**E, F**) in-plane magnetic hysteresis loops at different temperatures for both (**C, E**) pristine and (**D, F**) gated SRO films. In this measurement, the gated samples were obtained with *ex-situ* ILG at gating voltage of 3.5 V. The results show that a lightly reduced magnetism in the *ex-situ* gated sample due to the residual proton concentration within the film.

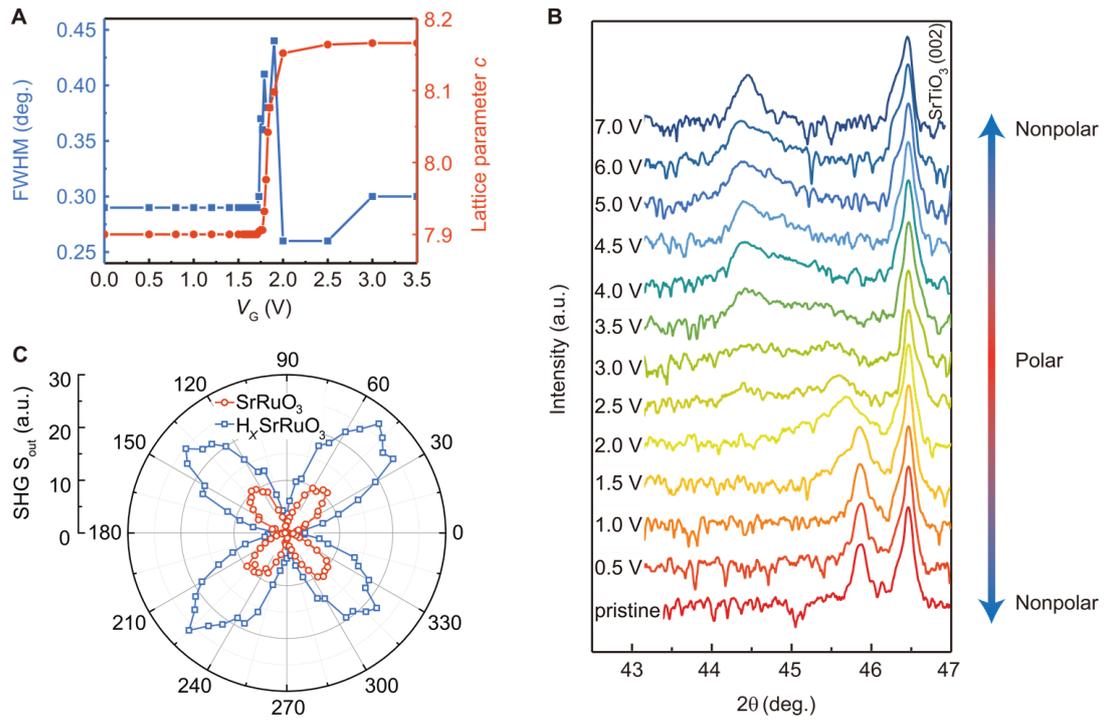

**Fig. S7. Emergence of a polar state during the ILG induced structural transformations.** (**A**) Correlation between the full width at half maximum (FWHM) of XRD $\theta$-$2\theta$ (002) diffraction peak of SrRuO$_3$ film (presented in **Fig. 1,** 28 nm) and the $c$-axis lattice constants as a function of $V_G$. (**B**) Detailed gating voltage dependent XRD $\theta$-$2\theta$ scans for a 90-nm SrRuO$_3$ thick film, in which a gradual structural-transformation is suggested based on the evolution of diffraction peaks. This result can be attributed to the formation of depth dependent proton concentration during the ILG, which is more pronounced at around the critical gating voltage. (**C**) Comparison of the $s$-polarized SHG intensity profiles for both pristine SrRuO$_3$ and gated H$_x$SrRuO$_3$ films. The enhanced SHG signal in the gated sample suggests the induced crystalline inversion symmetry with the gradient proton concentration.

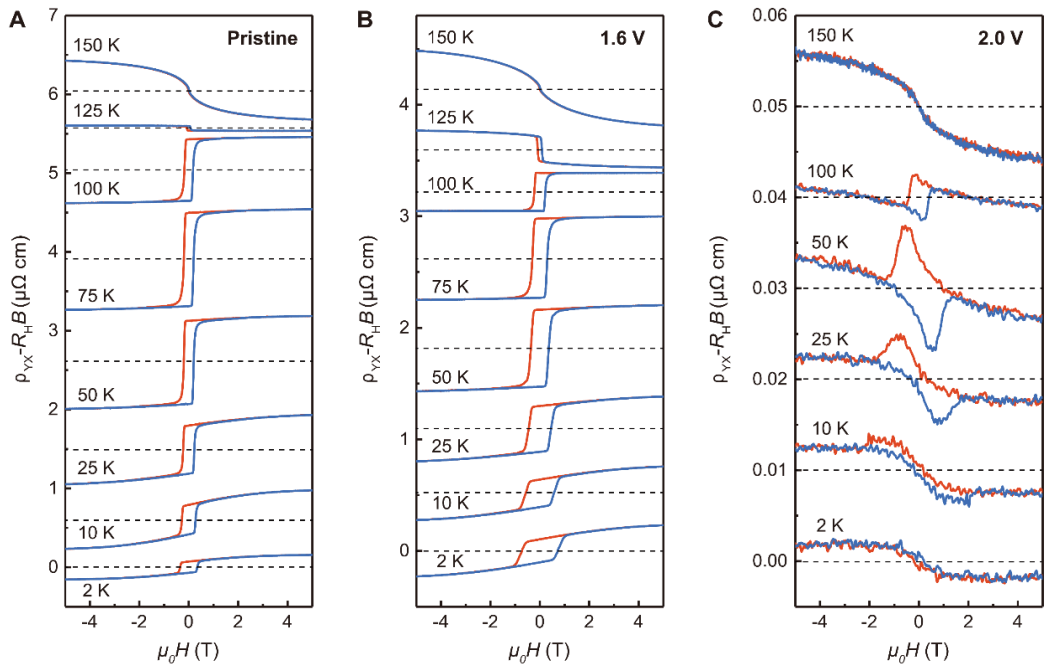

**Fig. S8. Temperature dependent anomalous Hall effect.** The pristine sample **(A)** and samples gated at **(B)** 1.6 V and **(C)** 2.0 V were used for the comparison. Ordinary Hall term was subtracted by the linear fits at higher magnetic field, and an offset is applied for each curve for clarity purpose.